# Emergent Language Symbolic Autoencoder (ELSA) with Weak Supervision to Model Hierarchical Brain Networks

Ammar Ahmed Pallikonda Latheef[1], Alberto Santamaria-Pang[2], Craig K Jones[1,3,4], Haris I Sair[3,4]

[1] Department of Computer Science, Johns Hopkins University, Baltimore MD 21218, USA
[2] Microsoft Health AI, Redmond Washington
[3] Department of Radiology and Radiological Science, Johns Hopkins School of Medicine, Baltimore MD 08544, USA
[4] Malone Center for Engineering in Healthcare, Johns Hopkins University, Baltimore MD 21218, USA

**Abstract.** Brain networks display a hierarchical organization, a complexity that poses a challenge for existing deep learning models, often structured as flat classifiers, leading to difficulties in interpretability and the 'black box' issue. To bridge this gap, we propose a novel architecture: a symbolic autoencoder informed by weak supervision and an Emergent Language (EL) framework. This model moves beyond traditional flat classifiers by producing hierarchical clusters and corresponding imagery, subsequently represented through symbolic sentences to improve the clinical interpretability of hierarchically organized data such as intrinsic brain networks, which can be characterized using resting-state fMRI images. Our innovation includes a generalized hierarchical loss function designed to ensure that both sentences and images accurately reflect the hierarchical structure of functional brain networks. This enables us to model functional brain networks from a broader perspective down to more granular details. Furthermore, we introduce a quantitative method to assess the hierarchical consistency of these symbolic representations. Our qualitative analyses show that our model successfully generates hierarchically organized, clinically interpretable images, a finding supported by our quantitative evaluations. We find that our best performing loss function leads to a hierarchical consistency of over 97% when identifying images corresponding to brain networks. This approach not only advances the interpretability of deep learning models in neuroimaging analysis but also represents a significant step towards modeling the intricate hierarchical nature of brain networks.

## 1 Introduction

Resting-state fMRI (rs-fMRI) has emerged as a key method for investigating brain connectivity, facilitating the study of the brain's functional networks. [5]. Traditional computational techniques, including General Linear Models (GLM) [11], Independent Component Analysis (ICA) [12], and Graph Theoretical approaches [3], have been used to analyze brain functional networks. However, these methods primarily focus on *segregating* brain networks without delving into the *interactions* between them [3,4,5]. Despite these advances, deep learning approaches encounter notable obstacles. One major challenge is their inherent lack of interpretability, often referred to as the 'black box' issue [6], which becomes particularly problematic in clinical contexts where understanding the rationale behind model predictions is essential [14, 23]. Furthermore, deep learning's extensive need for labeled data presents a significant hurdle, especially considering the limited availability of such data within the medical field [15,22].

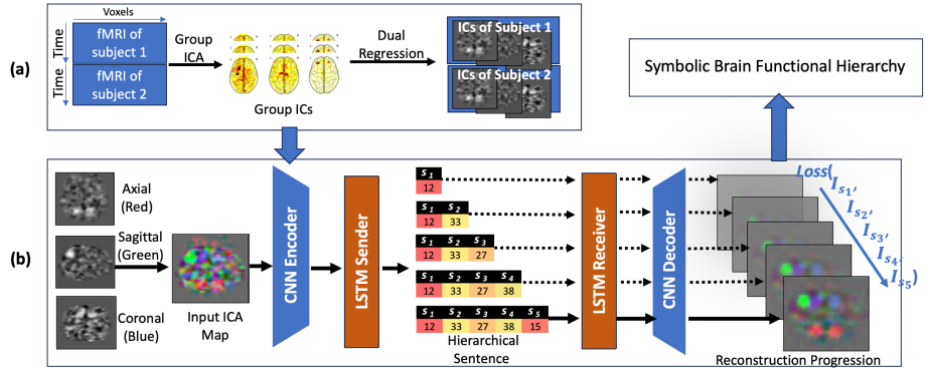

**Fig. 1.** (a) The Data processing module, (b) Model Architecture for ELSA contains an encoder, sender, receiver, and decoder. The dashed lines represent the modification to the architecture compared to previous works and are used for creating intermediate reconstructions. The set of reconstructions are used in our loss functions.

We introduce a groundbreaking approach, the Emergent Language Symbolic Autoencoder (ELSA), which utilizes a weakly supervised learning framework [8] integrated with Emergent Language (EL) [16,17,18,9,10] to produce hierarchical and progressive reconstructions of brain networks. ELSA's innovation lies in its progressive loss functions, specifically designed to enhance the hierarchical aspect of EL. These functions enable the model to generate sentences that reflect the transition from broad to specific representations of brain networks, significantly reducing the reliance on categorical labels during training. Unlike prior applications of EL, which lacked a mechanism for evaluating hierarchical representations, our method introduces an evaluation technique to ensure the sentences generated by ELSA maintain a coherent hierarchical structure. Additionally, ELSA is capable of generating interpretable, hierarchical sequences of resting-state brain function images, offering deep insights into the model's analytical process as it increasingly refines its understanding of network features.

Our key contributions are as follows: 1) We apply Emergent Language for hierarchical modeling of brain functional images, introducing a novel angle to brain network analysis; 2) We establish an innovative evaluation framework to assess the hierarchical integrity of sentences generated by our Emergent Language model; and 3)

We produce clear, hierarchically organized sequences of brain functional images, illustrating our model's progressively refined insight into network features.

We structure the paper as follows. In Section 2 we describe the dataset we use for analysis. Section 3 reviews previous methods and introduces our approach. Section 4 demonstrates qualitative and quantitative results. Section 5 we summarize our conclusion.

## 2      Dataset

Our study utilized the Beijing cohort dataset from the publicly accessible 1000 Functional Connectomes Project [25]. This dataset comprises resting-state fMRI (rs-fMRI) data from 198 young adults, including 76 males and 122 females, all aged between 18 and 26 years. The fMRI data for each participant consisted of 33 slices captured over 225 timepoints. We employed the FMRIB Software Library (FSL) [24] for data processing, adhering to established protocols for motion correction, spatial smoothing, temporal high-pass filtering, and aligning the data with the Montreal Neurological Institute (MNI) standard space. After processing, we performed a visual assessment of the data quality. This meticulous evaluation led to a refined dataset comprising 176 subjects [28].

To identify brain networks, we applied group Independent Component Analysis (ICA) [19], with the ICA order—representing the number of components—treated as a variable. We conducted five distinct runs of group ICA at varying component orders: 20, 40, 60, 80, and 100. To integrate the group ICA findings back into individual participant data, we used dual regression [20], a technique that projects the group ICA maps onto the individual fMRI data of each subject. For our analysis, the input ICA maps were generated by back-projecting these group ICA results onto the individual subject maps using dual regression, covering Axial, Sagittal, and Coronal views. These views were then combined into a single RGB image—Axial in red, Sagittal in green, and Coronal in blue—to produce composite input ICA maps that visually represent the brain networks.

## 3      Emergent Language Symbolic Autoencoder (ELSA)

Previous research on Emergent Language (EL) introduced Symbolic Semantic Segmentation and Symbolic Variational Autoencoders [10]. These approaches utilize sender-receiver framework to produce symbolic sentences. These sentences contain semantic details of segmentation or image content. In our current work, we aim to further explore the potential of EL by focusing on the semantic and hierarchical structures inherent in brain networks. Specifically, we seek to create semantic images that accurately reflect the learned hierarchical organization of resting-state brain networks, thereby extending the interpretability and application of EL in neuroradiology.

We formalize ELSA as the process of integrating symbolic sequences with corresponding visual representations through a hierarchical generative model. Let the vocabulary, $\Sigma = \{s_1, s_2, ..., s_n\}$ denote a set of symbols from which sequences are formed. A sequence $S = (s_1, s_2, ..., s_k)$, with $s_i \in \Sigma$, $1 \leq i \leq k$ and $k \leq n$ represents a specific arrangement of symbols. The task for the generative model involves producing a mapping $\mathcal{M}: S \times N \rightarrow I$, where $S$ is a sequence of symbols, $N$ is the natural numbers denoting the sequence's length, and $I$ represents the space of generated images. For a given sequence $S$, the mapping is defined as:

1. For each prefix of $S$, denoted $S = (s_1, ..., s_i)$, there exists a corresponding image $I_{S_i}$ that visually encapsulates the information contained in $S_i$.
2. The mapping $\mathcal{M}$ thus generates a set of pairs $\{(S_i, I_{S_i})\}_{i=1}^{k}$, , where each pair consists of a prefix of the sequence $S$ and its associated image $I_{S_i}$.

This model requires the generative network to not only recognize and encode the symbolic information within $S$ but also to iteratively construct a visual representation for each incremental addition to the sequence. The final output is a collection of images $\{I_{S_i}\}_{i=1}^{k}$ that represent the hierarchical structure embedded within the sequence $S$, showcasing the network's capability to translate symbolic data into coherent visual signals. The loss function is estimated from the collection of images $\{I_{S_i}\}_{i=1}^{k}$ which represent a hierarchical structure.

The mathematical description of the CNN-Based Encoder-Decoder architecture within the Emergent Language (EL) framework is detailed as follows:

**CNN-Based Encoder:** Let us define $P = \{I_{S_i}\}_{i=1}^{k}$ as the sequence of images corresponding to different Independent Component Analysis (ICA) component maps. The encoder is composed of four Convolutional Neural Network (CNN) layers [21] followed by two linear layers, mathematically represented by: Encoded = $L_2 \circ CNN_4 \circ ... \circ CNN_1(P)$ where $CNN_j$ is the $j^{th}$ CNN layer and $L_i$ is the $i^{th}$ linear layer. This structure distills the features from $P$ into a compact latent representation.

**Transmission to EL Encoder (Sender):** This encoded representation is then passed to the EL encoder (sender), transforming it into a sequence of discrete symbols, represented as EL_Encoded = Sender(Encoded). The sender network converts the CNN-encoded ICA maps into sequences of discrete categorical variables [16][17], forming symbolic sentences that encapsulate hierarchical information [8][10].

**EL Decoder (Receiver) Processing:** Diverging from models where the receiver only processes the final hidden state [18], our receiver decodes every hidden state from the sender, allowing for iterative image reconstruction at each symbol in the sentence. For each symbol $s_i$ and its corresponding hidden state $h_i$, the decoder produces a partial reconstruction $R_i$, iterating over the sentence length $N$: $R_i = \text{Decoder}(h_i)$,    $\forall i \in$ $\{1, \dots, N\}$. This iterative decoding permits the reconstruction of multiple images, each correlating to the cumulative sequence of symbols.

**CNN-Based Decoder:** The output from the EL decoder is then processed through a CNN-based decoder, comprising two linear layers and four CNN layers, tasked with the final reconstruction of the ICA map. This process can be denoted as: Reconstructed ICA Map = $CNN_4^{-1} \circ \dots \circ CNN_1^{-1} \circ L_2^{-1}(\text{EL\_Decoded})$. Here, $CNN_j^{-1}$ and $L_i^{-1}$ represent the inverse operations of the CNN and linear layers in the decoder, respectively, with EL_Decoded as the output from the EL decoder.

We introduce a general framework allows us to specify the type of loss calculation based on the model's current training objective and the hierarchical nature of the data being processed. We define a hierarchical generalized loss function $L_h$ that can adapt based on two key parameters: $\alpha$ and $\beta$. These parameters control the scope of the loss computation (over the entire sequence or progressively) and the strictness of the evaluation (considering each output versus the final output). Given a sequence $S = (s_1, s_2, \dots, s_k)$ and its corresponding generated images $\{I_{S_i}\}_{i=1}^{k}$, and the original input image $X$, the generalized loss function can be represented as:

$$L_h(X, \{I_{S_i}\}; \alpha, \beta) = L_{cb}(X, I_{S_{\alpha k}}) + \sum_{i=1}^{\alpha k - 1} \beta \cdot L_{cb}(X, I_{S_i}) \qquad (1)$$

Where, $\alpha$ is a parameter that determines the coverage of the sequence. For $\alpha = 1$, the loss is calculated over the entire sequence. For $0 < \alpha < 1$, the loss calculation is progressively limited to the first $\alpha k$ terms of the sequence, where $k$ is the total length of the sequence. $\beta$ is a binary weight that decides whether the loss function considers the first $\alpha k - 1$ terms.

The term $L_{cb}$ is a containing bias term to encourage the model to create broader representations when suitable, i.e., in the initial symbols of a sentence. We apply $w_{large}$ weight when a pixel in the output image is larger than the corresponding pixel in the input image, and $w_{small}$ when a corresponding pixel in the output image is smaller than the corresponding pixel in the input image. Given input image $X$ and an output image $I$, both flattened and consisting of N pixels, the containing bias loss, $L_{cb}$ is expressed as:

$$L_{cb}(X, I) = (1/N) \begin{cases} \sum_{i=1}^{N} w_{small}(I_i - X_i)^2, & I_i < X_i \\ \sum_{i=1}^{N} w_{large}(I_i - X_i)^2, & I_i \geq X_i \end{cases} \qquad (2)$$

## 4    Experiments and Results

The hierarchical generalized loss function accommodates various training objectives and data hierarchies by adjusting $\alpha$ and $\beta$. We apply the "regular" loss function used in previous works by setting $\alpha = 1$ and $\beta = 0$. We create the "strict" loss function by setting $\alpha = 1$ and $\beta = 1$. We create the "progressive" loss function by setting $\alpha$ according to the ICA order of input image and $\beta = 0$. Finally, our "progressive strict" loss function is created by setting $\alpha$ according to input image ICA order and $\beta = 1$.

| | Symbol 1 | Symbol 2 | Symbol 3 | Symbol 4 | Symbol 5 | Brain Network | ICA Order |
|---|---|---|---|---|---|---|---|
| Sample 1 | 12 | 33 | 27 | 38 | 15 | DMN-IPL-Left | 100 |
| Sample 2 | 12 | 33 | 27 | 38 | 89 | DMN-IPL-Left | 100 |
| Sample 3 | 12 | 33 | 27 | 74 | 26 | DMN-IPL-Left | 80 |
| Sample 4 | 12 | 33 | 27 | 74 | 46 | DMN-IPL-Left | 80 |
| Sample 5 | 12 | 33 | 92 | 34 | 81 | DMN-IPL-Right | 80 |
| Sample 6 | 12 | 33 | 92 | 34 | 97 | DMN-IPL-Right | 80 |
| Sample 7 | 12 | 33 | 92 | 58 | 99 | DMN-IPL-Right | 60 |
| Sample 8 | 12 | 33 | 38 | 44 | 82 | DMN-IPL | 40 |
| Sample 9 | 12 | 49 | 51 | 79 | 14 | DMN-RSC | 40 |
| Sample 10 | 12 | 49 | 51 | 79 | 31 | DMN-RSC | 40 |
| Sample 11 | 12 | 59 | 12 | 37 | 90 | DMN | 20 |
| Sample 12 | 18 | 39 | 28 | 17 | 58 | Motor-Dorsal-Leg | 60 |
| *Sample n* | *n1* | *n2* | *n3* | *n4* | *n5* | | |

**Fig. 2.** Figure 2 illustrates a sample table that captures the sentences produced by the ELSA model's sender for each data point in our study.

We use our model to generate hierarchical sentences, which can be systematically organized. They represent input ICA maps and vary in length according to the granularity of the group ICA orders applied to the fMRI data (20, 40, 60, 80, 100). With increasing ICA orders, the components represent more specific brain network subcomponents or functions. The hierarchical nature means that sentences of different lengths correspond to ICA maps of different orders. The capability of ELSA to generate hierarchical sentences mirrors the granularity inherent in the input ICA maps, informed by the group ICA orders (20, 40, 60, 80, 100) applied to the fMRI data (detailed in Section 2). These ICA orders establish a tiered model that unveils increasingly precise features of brain network components or functions. For example, a broader correlation among brain networks is observed with 20 independent components, whereas more detailed representations emerge at 100 independent components. This tiered approach results in sentences of variable lengths that correspond to distinct ICA orders. As illustrated in Figure 2, segments of each sentence are highlighted to signify their relevance to the represented ICA order, with the unhighlighted portions considered as expansions by the model. Each dataset entry correlates with a unique ICA component derived from individual participants, with the model crafting a specific five-word sentence for each analyzed ICA volume. For instance, the Default Mode Network (DMN) is represented by the sentence (12), and its association with the Inferior Parietal Lobule (DMN-IPL) is articulated through the sentence (12, 33). Notably, in the context of an ICA order of 100, all sentence symbols are accentuated, whereas for an ICA order of 80, only 80% of the symbols (4 out of 5) are emphasized, as demonstrated in Figure 2.

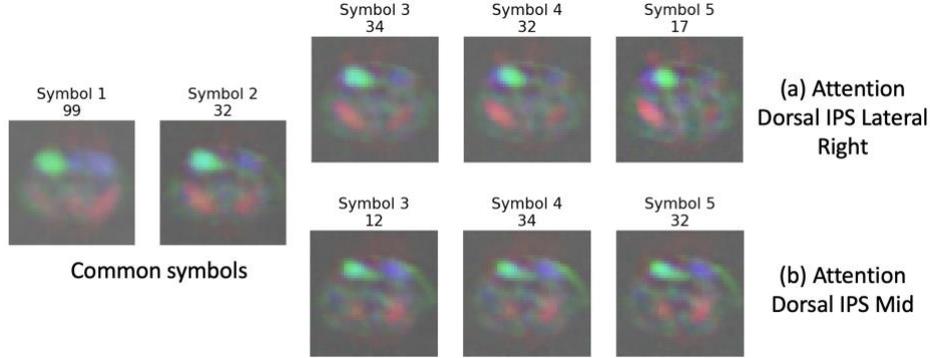

**Fig. 3.** Progressive reconstruction of ICA images based on sentence symbols. Sentences (a) and (b) initially share symbols, diverging at the third symbol.

Past works in EL did not evaluate the hierarchical consistency within generated sentences. To measure hierarchical consistency at a given ICA order (or sentence length as defined in fig 2), we calculate the percentage of sentences of the given ICA order with shared common prefix (of length defined by the given ICA order) with sentences of ICA order 100, that also have the same parent network. Thus, we calculate an accuracy value for each ICA order.

**Qualitative Results:** Figure 3 shows the progressive reconstructions of two ICA input images, the first is an Attention Dorsal IPS Lateral Right network and the second is an Attention Dorsal IPS Mid network using the progressive strict with containing bias loss function. The result shows the capability of our model to create interpretable, hierarchical progressions of images from their parent-like networks in the initial symbols to the specific network by the final symbol. But we do observe that from symbol 1 to symbol 2, the activations do not become more specific, instead they shift. Suboptimal reconstructions for symbol 1 were common across various images, indicating one of the limitations of our work. Figure 4 shows the effect of our newly introduced loss functions. Note that the regular loss function does not create sequences that can be interpreted as a hierarchical progression from parent-like to the specific network. Utilizing the strict loss function yields interpretable progression, however it does not seem hierarchical. On the other hand, our progressive strict with containing bias loss function does give such a hierarchical sequence of images, going from broad parent-like representations to more specific representations.

**Quantitative Results:** Table 1 indicates that progressive strict with containing bias performs the best at lower ICA orders but performs similar or worse compared to progressive strict at higher orders. The table shows that using the "strict" constraint with the (regular) progressive loss function improves accuracy a lot. All models perform badly at ICA 20, which matches the bad representations at symbol 1 in Figure 4.

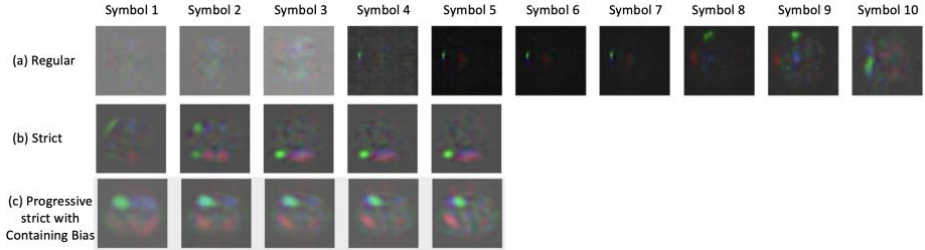

**Figure 4:** Example of different image hierarchical reconstruction. Reconstruction of DMN-Parahippocampal in (a), Cerebellar Superior Mid Posterior in (b), and Attention Dorsal IPS Lateral Right in (c).

The model achieves perfect accuracy with a sentence length of 10 at both ICA 80 and ICA 100. In contrast, a sentence length of 5 attains near-perfect accuracy only at ICA 100. This suggests that longer sentences enable the model to capture network patterns more precisely. The Progressive Strict with Containing Bias condition excels at lower ICA orders (20 and 40), highlighting its effectiveness in the early stages of component analysis. Across all loss function conditions, a sentence length of 10 consistently delivers higher accuracies compared to a sentence length of 5.

**Table 1.** Hierarchical quality of sentences at different ICA orders. Loss functions used include progressive, progressive Strict (S) and progressive strict with Containing Bias (CB).

| Sentence Length | Loss Function | Quality of Hierarchy | | | | |
| --- | --- | --- | --- | --- | --- | --- |
| | | @ ICA 20 | @ ICA 40 | @ ICA 60 | @ ICA 80 | @ICA 100 |
| **10** | Progressive | 28.1 | 50.7 | 94.1 | **100.0** | **100.0** |
| | Progressive S | 32.8 | 62.7 | 97.1 | **100.0** | **100.0** |
| | Progressive S CB | **43.5** | **75.0** | **97.6** | **100.0** | **100.0** |
| **5** | Progressive | 20.8 | 37.6 | 61.7 | 87.5 | 99.93 |
| | Progressive S | 26.7 | 49.7 | **84.7** | **98.2** | **99.99** |
| | Progressive S CB | **29.8** | **51.1** | 80.9 | 97.3 | 99.93 |

## 5   Conclusion

Our study presents an innovative method for modeling hierarchical brain networks using a weakly supervised deep learning framework, enriched with Emergent Language (EL). Through the introduction of progressive loss functions and a novel evaluation method for assessing hierarchical consistency, we have successfully applied EL to hierarchically structured medical imaging data. Our findings reveal that these progressive loss functions result in more interpretable hierarchical reconstructions than those achieved with traditional loss functions.